\documentstyle[psfig,aps,prl,multicol]{revtex}

\begin{document}

\title{Negative entropy and information in quantum mechanics}
\author{N. J. Cerf$^1$ and C. Adami$^{1,2}$}
\address{$^1$W. K. Kellogg Radiation Laboratory and 
         $^2$Computation and Neural Systems\\
California Institute of Technology,
Pasadena, California 91125, USA}

\date{December 1995}

\maketitle
\draft

\begin{abstract}
A framework for a quantum mechanical information theory
is introduced that is based entirely on density operators,
and gives rise to a unified description of classical correlation
and quantum entanglement. Unlike in classical (Shannon) information theory, 
quantum (von Neumann) conditional entropies can be
negative when considering quantum entangled systems, a fact related
to quantum non-separability. 
The possibility that negative (virtual) information can be carried
by entangled particles suggests a consistent interpretation
of quantum informational processes.
\end{abstract}
\bigskip
\pacs{PACS numbers: 03.65.Bz,05.30.-d,89.70.+c
      \hfill KRL preprint MAP-191}

\begin{multicols}{2}[]
\narrowtext

Quantum information theory is a new field with potential implications
for the conceptual foundations of quantum mechanics. It appears to be
the basis for a proper understanding of the emerging fields of quantum
computation \cite{bib_quantcomp}, quantum communication, and quantum
cryptography \cite{bib_crypto}.  Although fundamental results 
on the quantum noiseless coding theorem
\cite{bib_schumacher} and the capacity of quantum noisy channels
\cite{bib_extraction} have been obtained recently, quantum information
is still puzzling in many respects. This is especially true for 
quantum teleportation and superdense coding,
two purely quantum communication schemes devised recently \cite{bib_teleport}.
Indeed, these dual processes
which rely on the quantum correlation between the two members of a
spatially separated Einstein-Podolsky-Rosen (EPR) pair are difficult
to interpret in terms of information theory.  
We show in this Letter that these processes can be understood in a
consistent way by exploiting a fundamental difference between
Shannon theory~\cite{bib_shannon} and an extended information
theory that accounts for quantum entanglement.
As we shall see, the latter allows for {\em negative}
conditional entropy even though this is forbidden classically. 
This leads us to propose that such quantum informational 
processes can be described by diagrams---much like particle physics
reactions---involving particles carrying {\em negative} (virtual)
information. By analogy with anti-particles, we refer to them as
{\em anti-qubits}.
\par

Previous attempts to describe quantum informational processes have generally
relied on the formulas of classical information theory supplemented with
quantum probabilities, {\em not} amplitudes. However, it has
been realized since Schumacher~\cite{bib_schumacher} that the
von Neumann entropy has an {\em information-theoretical} meaning,
characterizing (asymptotically) the minimum amount of quantum
resources required to code an ensemble of quantum
states. This suggests that an {\em extended} information
theory can be defined that explicitly takes quantum phases into account,
as attempted in this Letter. The theory described here
characterizes multipartite quantum systems
using only density operators and von Neumann entropies.
It includes Shannon theory as a special case
but describes quantum entanglement as well, thereby providing
a {\em unified} treatment of classical and quantum information.
To be specific, let us consider a composite system consisting of two
(classical or quantum) variables, 
$A$ and $B$, and outline in parallel the classical and our quantum 
information-theoretic treatment of it. 
In classical information theory, 
we define the Shannon entropy of $A$~\cite{bib_shannon},
\begin{equation}
H(A)=-\sum_a p(a) \log_2 p(a)  \;,
\end{equation}
where the variable $A$ takes on value $a$ with probability $p(a)$.
It is interpreted as the uncertainty about $A$ 
[an analogous definition holds for $H(B)$]. The quantum analog
is the von Neumann entropy $S(\rho_A)$ of a quantum source $A$
described by the density operator $\rho_A$,
\begin{equation}
S(A)=-{\rm Tr}_A [\rho_A \log_2 \rho_A]  \;,
\end{equation}
where ${\rm Tr}_A$ denotes the trace over the degrees of freedom
associated with $A$. The von Neumann entropy reduces to a Shannon entropy
if $\rho_A$ is a mixed state composed of orthogonal quantum states.
The combined classical system $AB$ is characterized by a joint probability
$p(a,b)$, and therefore by a joint entropy
$H(AB)=-\sum_{a,b} p(a,b) \log_2 p(a,b)$
which is the uncertainty about the entire system. The quantum definition
$S(AB)=-{\rm Tr}[\rho_{AB}\log_2 \rho_{AB}]$,
a function of the density operator of the combined system $\rho_{AB}$,
is immediate. The classical probabilities observe $p(a)=\sum_b p(a,b)$ 
and $p(b)=\sum_a p(a,b)$; analogously,
$\rho_A={\rm Tr}_B[\rho_{AB}]$ and $\rho_B={\rm Tr}_A[\rho_{AB}]$.
Note that ${\rm Tr}_A$ and ${\rm Tr}_B$ stand for partial traces,
while ${\rm Tr}$ is the trace over the joint Hilbert space.
The classical {\em conditional} entropy is defined as
$H(A|B)=H(AB)-H(B)=-\sum_{a,b} p(a,b) \log_2 p(a|b)$,
where $p(a|b)=p(a,b)/p(b)$ is the probability of $a$ conditional on $b$;
$H(A|B)$ characterizes the remaining uncertainty about the source $A$
when $B$ is known.
\par

Here, we propose that the correspondence between classical and
quantum constructions can be extended 
by defining the von Neumann {\em conditional} entropy
\begin{equation}  \label{eq_cond_entr}
S(A|B) = - {\rm Tr}[\rho_{AB} \log_2 \rho_{A|B}]
\end{equation}
based on a conditional ``amplitude'' operator
\begin{equation}  \label{eq_cond_dens_matr}
\rho_{A|B}= \exp_2 (-\sigma_{AB})
= \lim_{n \to \infty}
\left[ \rho_{AB}^{1/n} ({\bf 1}_A \otimes \rho_B)^{-1/n} \right]^n  ,
\end{equation}
where $\sigma_{AB}={\bf 1}_A \otimes \log_2 \rho_B-\log_2 \rho_{AB}$,
${\bf 1}_A$ being the unit matrix in the Hilbert space of
$A$, and $\otimes$ the tensor product 
in the joint Hilbert space. Eq. (\ref{eq_cond_dens_matr})
is a quantum generalization of the conditional probability $p(a|b)$
and reduces to it in the classical limit (diagonal $\rho_{AB}$)~\cite{fn_math}.
In general, $\rho_{A|B}$ is a positive Hermitian operator
on the joint Hilbert space [just as $p(a|b)$ is a function of $a$ and $b$],
whose spectrum is invariant under frame changes of the product
form $U_A \otimes U_B$ on $\rho_{AB}$. 
We refer to $\rho_{A|B}$ as an {\em amplitude} operator
to emphasize that it retains the quantum phases, in contrast to $p(a|b)$. 
It is defined on the support of $\rho_{AB}$ since the latter is
included in the support of ${\bf 1}_A \otimes \rho_B$, so that
$S(A|B)$ is well-defined. Indeed, inserting 
Eq.~(\ref{eq_cond_dens_matr})
in Eq.~(\ref{eq_cond_entr}) results in $S(A|B)=S(AB)-S(B)$,
just as for Shannon entropies.
However, $\rho_{A|B}$ is {\em not} a density operator as its
eigenvalues can exceed 1, in which case the operator $\sigma_{AB}$
is {\em not} positive semi-definite, in contrast to its classical counterpart.
It is precisely for this reason that the von Neumann {\em conditional}
entropy can be {\em negative}.
In Shannon theory, $H(A|B)$ is always non-negative, reflecting
that the classical entropy
of a composite system $AB$ cannot be lower than the entropy of
any subsystem $A$ or $B$, i.e., $\max [ H(A),H(B) ] \le H(AB)$.
Quantum entropies on the other hand are known to be non-monotonic, i.e.,
$S(A) > S(AB)$ or $S(B) > S(AB)$ is possible when 
$A$ and $B$ are quantum entangled subsystems (see, e.g., \cite{bib_wehrl}).
The above operator-based formalism naturally accounts 
for this non-monotonicity.
\par

The appearance of nonclassical ($>1$) eigenvalues of $\rho_{A|B}$
and nonclassical ($<0$) conditional entropies
can be related to quantum non-separability and the
violation of entropic Bell inequalities, 
as we show in~\cite{bib_bellpaper}. It can be proven that
any separable state 
$\rho_{AB}= \sum_i w_i \rho_A^{(i)} \otimes \rho_B^{(i)}$ 
(with $0\le w_i \le 1$ and $\sum_i w_i =1$) is such that
$\sigma_{AB}\ge 0$, and is therefore
associated with $S(A|B)={\rm Tr}[\rho_{AB}\,\sigma_{AB}]\ge0$
\cite{fn_separ}. Thus, {\em all} the eigenvalues of
$\rho_{A|B}$ (and $\rho_{B|A}$) are $\le 1$ for any convex
mixture of product states, which implies that it is a
{\em necessary} condition for separability
in a Hilbert space of arbitrary dimensions. 
A negative conditional entropy implies that $\rho_{A|B}\not\le 1$ while
the converse is not true, so that
$S(A|B)\ge 0$ is a weaker necessary condition. As an example,
this criterion can be applied to two spin-1/2 particles
in a Werner state, i.e., a mixture
of a singlet fraction $x$ and a random fraction $(1-x)$.
A simple calculation
shows that $\rho_{A|B}$ admits three eigenvalues equal to $(1-x)/2$, and
a fourth equal to $(1+3x)/2$. 
The above separability criterion is thus fulfilled
when the latter does not exceed 1, that is for $x \le 1/3$.
Thus, for this particular case (in fact, for any mixture of Bell states),
our condition is simply equal to Peres'~\cite{bib_peres} and then happens
to be also sufficient.
Numerical evidence suggests, however, that it is weaker than Peres'
in general. This will be investigated elsewhere.
\par

In Shannon information theory, we define
the {\em mutual} (or correlation) entropy $H(A{\rm :}B)=H(A)-H(A|B)$
as the decrease of the entropy of $A$ due to the knowledge of $B$, 
resulting in
$H(A{\rm :}B)=H(A)+H(B)-H(AB)=-\sum_{a,b} p(a,b) \log_2 p(a{\rm :}b)$,
where $p(a{\rm :}b)=p(a)p(b)/p(a,b)$ is a {\em mutual} probability.
The mutual entropy corresponds to the
{\em information} gained about $A$ by measuring $B$. It is symmetric,
i.e., $H(A{\rm :}B)=H(B{\rm :}A)$, and can be viewed
as the amount of entropy shared by $A$ and $B$. 
Also, $H(A{\rm :}B) \ge 0$,
as the entropy of $A$ can only be {\em reduced}
through the knowledge of $B$. 
As we now show, a quantum mechanical extension of this concept
can be straightforwardly defined, i.e., we can construct 
a von Neumann {\em mutual} entropy $S(A{\rm :}B)$ based on
a {\em mutual} amplitude operator
\begin{equation}  \label{eq_mutual}
\rho_{A:B}=\lim_{n\to\infty}
\left[ (\rho_A \otimes \rho_B)^{1/n} \rho_{AB}^{-1/n} \right]^n  \;,
\end{equation}
generalizing the mutual probability $p(a{\rm :}b)$. 
Again, the associated quantum {\em mutual} entropy
\begin{equation}
S(A{\rm :}B) = -{\rm Tr}[\rho_{AB} \log_2 \rho_{A:B}]
\end{equation}
is invariant under frame changes of the product form,
and can be written as $S(A{\rm :}B)=S(A)+S(B)-S(AB)$.
The quantum mutual entropy $S(A{\rm :}B)$ is a natural extension of
$H(A{\rm:}B)$ which measures quantum as well as classical correlations,
and reduces to it for diagonal density matrices
(probability distributions). It does {\em not} discriminate
purely quantum entanglement from correlation, however, but rather unifies
their information-theoretic description.
While $S(A{\rm :}B)\ge 0$ just as its classical counterpart,
it can exceed the entire uncertainty of the source ensemble, namely
\begin{equation}  \label{eq_quant_mutual}
S(A{\rm :}B) \le 2 \min [S(A),S(B)] \; ,
\end{equation}
as implied by the Araki-Lieb inequality (see~\cite{bib_wehrl}).
This precisely occurs for quantum entangled subsystems,
and is forbidden classically due to the non-negativity
of the Shannon conditional entropy, i.e.,
$H(A{\rm :}B) \le \min [H(A),H(B)]$.
\par

\begin{figure}
\caption{ (a) Entropy Venn diagram for a bipartite system $AB$. 
(b) Diagram for two qubits
with $S(A)=S(B)=1$: 
(I)~independent 50/50 mixtures of states $|0\rangle$ and $|1\rangle$;
(II)~maximally classically (anti)correlated qubits,
{\it i.e.}, a 50/50 mixture
of $|01\rangle$ and $|10\rangle$;
(III)~fully entangled EPR state with wavefunction
$2^{-1/2} (|01\rangle - |10\rangle)$,
or any Bell state in general.  }
\label {fig_venn}
\vskip 0.5cm
\centerline{\psfig{figure=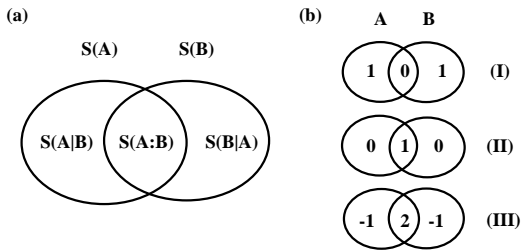,width=2.70in,angle=0}}
\vskip -0.2cm
\end{figure}

The above unified treatment of classical and quantum correlation
is illustrated by considering three limiting cases
and their associated entropy Venn diagrams 
as defined in Fig.~\ref{fig_venn}a. For independent
quantum systems (I), one has $\rho_{AB}=\rho_A \otimes \rho_B$,
so that $\rho_{A|B}=\rho_A \otimes {\bf 1}_B$ [analogous to $p(a|b)=p(a)$]
and $S(A|B)=S(A)$, thereby saturating the lower bound $S(A{\rm:}B)=0$
as in the classical case. This is illustrated in Fig.~\ref{fig_venn}b for 
two quantum bits (qubits), i.e., if systems $A$ and $B$ 
belong to a 2-state Hilbert space.
For two (maximally) classically correlated systems (II),
the {\em classical} upper bound $S(A{\rm:}B)=\min[S(A),S(B)]$ 
is saturated.
The range between the classical and quantum upper bounds
corresponds to a purely quantum (classically {\em forbidden}) regime,
namely quantum entanglement. The {\em quantum} upper bound,
Eq.~(\ref{eq_quant_mutual}),
is saturated for maximally entangled systems such as EPR pairs (III):
the conditional entropies are negative, while
$S(A{\rm :}B)$ exceeds the value which defines 100\% correlations,
so that entanglement might be viewed as {\em supercorrelation}. 
\par

\begin{figure}
\caption{ Physical spacetime diagrams
and quantum information dynamics diagrams
for (a) quantum teleportation and (b) superdense coding.}
\label{fig_diagrams}
\vskip 0.3cm
\par
%\centerline{\psfig{figure=fig2.ps,width=2.70in,angle=0}}
\centerline{\psfig{figure=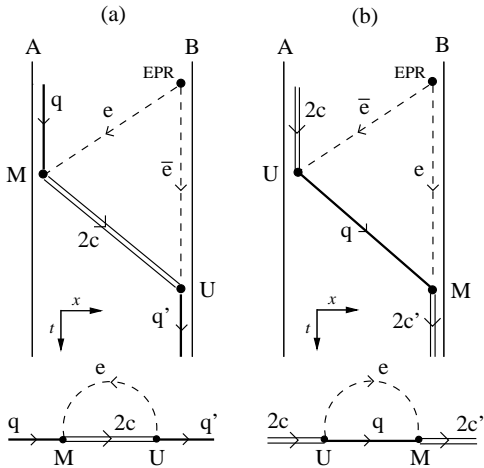,width=2.50in,angle=-90}}
\vskip -0.2cm
\par
\end{figure}

The information-theoretic formalism discussed here should thus be regarded
as an {\em extension} of a classical formalism beyond its original
range to identify the signature of quantum dynamics.
As shown in further work (see \cite{bib_physicad} and references therein), 
it can be successfully applied to {\em multipartite} entangled systems
by extending Shannon construction
[e.g., defining the quantum conditional mutual entropy $S(A{\rm:}B|C)$].
Most of the standard
concepts of Shannon theory then find an intuitive quantum analog, which
results in a convenient framework for analyzing quantum channels or
error correction, for example. Here, we restrict ourselves 
to sketching an information-theoretic description of
quantum teleportation and superdense coding 
relying on this framework.
We argue that a conservative picture of information flow in these processes,
consistent with unitarity, is only possible if the existence 
of negative conditional entropies is recognized~\cite{fn_elsewhere}. 
\par

The third diagram in Fig.~\ref{fig_venn}b exhibits how the negative 
conditional entropy of a member of an EPR pair, $S(A|B)=-1$,
is balanced by the (unconditional) entropy $S(B)=1$ of its partner.
As entropy can be viewed as ``latent'' information,
this suggests characterizing the members of an EPR pair
by a {\it virtual}
information content of $\pm1$ bit~\cite{virtualinfo},
a concept that turns out to be very fruitful when analyzing the
information flow.
To distinguish these maximally entangled qubits,
we call them {\em ebits}~\cite{ebit}. In this 
picture inspired by particle physics, 
the ebit ($e$) and anti-ebit ($\overline{e}$) can be viewed as virtual
conjugate particles, and the preparation of an EPR
pair appears as the creation of an ``$e\overline{e}$ pair'' from the
entropy ``vacuum'' $S(e\overline{e})=0$. The central point is
that the virtual information content of an
ebit can be revealed (i.e., converted to real information) in
interactions with qubits and classical bits (cbits), as we shall see.
In quantum teleportation (see Fig.~\ref{fig_diagrams}a),
an unknown qubit ($q$) is transported with
perfect fidelity through the transmission of two cbits ($2c$), after
the sender and the receiver have shared an $e\overline{e}$ pair. The
sender performs a joint measurement (M) of the qubit and the ebit in
the 2-particle Bell basis (i.e., 4 orthogonal maximally entangled
2-particle states), thereby generating 2 cbits. The receiver
reconstructs the qubit from the 2 cbits by applying to
$\overline{e}$ one of 4 possible unitary transforms (U) in the
1-particle Hilbert space.  Only if the (virtual)
information content of $e$ and
$\overline{e}$ is taken into account properly (1~bit and --1~bit,
respectively) can the information flow be conserved through the
(M) and (U) stages.
More specifically, if $q$ is maximally entangled with an external
reference (not represented here), the entropy conservation rule for (M) is
\begin{equation}   \label{eq_1/4}
S(2c)=S(qe)=S(q)+S(e)=1+1=2     \;,
\end{equation}
since $q$ and $e$ are initially independent,
while the {\em conditional} entropy
of the remaining particle is $S(\overline{e}|qe)=-1$.
At (U), we have
\begin{equation}
S(q')=S(qe\overline{e})=S(qe)+S(\overline{e}|qe)=2-1=1  \;,
\end{equation}
where $q'$ is the outgoing qubit~\cite{fn_leftover}.
In the particle physics language,
if the $\overline{e}$ is replaced by an $e$ going backwards in time (see
diagram in lower half of Fig.~\ref{fig_diagrams}a), the whole process
is formally equivalent to the transmission of 1 qubit via 2 cbits, but
with the additional burden of 1 ebit. A violation of causality by the
$\overline{e}$ is prevented by the fact that its information content
cannot be revealed without the presence of the 2 cbits, which travel
causally.  
\par

In superdense coding (see Fig.~\ref{fig_diagrams}b), 
2 cbits are apparently transported via 1 qubit (a 2-state particle!).
Our analysis suggests that the negative information content of the
$\overline{e}$ is exploited by the sender so that 2 cbits can be
packed into a single qubit via the unitary transform (U).
Subsequently, the receiver performs a joint measurement (M) of the
qubit and ebit in the 2-particle Bell basis, thereby recovering the
two encoded cbits. The entropy conservation rule for (U) is
\begin{equation}
S(q|e)=S(2c\;\overline{e}|e)=S(2c)+S(\overline{e}|e)=2-1=1  \;,
\end{equation}
since $2c$ and $\overline{e}$ are initially independent,
while the {\em unconditional} entropy of the remaining particle is
$S(e)=1$. At (M), we have 
\begin{equation}   \label{eq_4/4}
S(2c')=S(qe)=S(q|e)+S(e)=1+1=2  \;,
\end{equation} 
where $2c'$ are the outgoing cbits. The cbits $2c$ left with the sender
must be ignored
from an informational point of view, as they are correlated with $2c'$.
The factor 2 that is {\em apparently} gained here is directly connected to
the factor 2 in Eq.~(\ref{eq_quant_mutual}), and originates from the
fact that $S(2c{\rm:}q|e)=2$, that is, the additional information (about
$2c$) conveyed by $q$ when knowing $e$ exceeds the entropy of $q$ by a
factor of 2. It is the maximum compression 
allowed by quantum mechanics (no other quantum communication scheme
could be more efficient).  However, this compression is only {\em
apparent} since the information flow is conserved in both (U) and (M)
stages if a qubit--antiqubit picture is
used.  Formally, the 2 cbits are distributed over {\em two}
particles (see lower half of Fig.~\ref{fig_diagrams}b) although the
ebit does not appear during the considered period of transmission as
it is sent backwards in time. 
\par

We have proposed a quantum mechanical extension of Shannon information
theory which incorporates entanglement via negative conditional entropies.
This analysis suggests the possibility
that a qubit (the fundamental quantum of information) could have
an analogously defined {\em anti-qubit} (a quantum of {\em negative}
information),
formally equivalent to a qubit traveling backwards in time as anticipated
in \cite{bib_teleport}. 
The concept that negative virtual information
can be carried by entangled particles provides interesting insight
into the information flow in quantum communication
processes such as quantum teleportation and superdense coding.
Furthermore, this leads us to conjecture that these processes
can be recast into reactions involving information quanta
($c$, $q$, $e$, $\overline{e}$)
described by {\em diagrams}, much like particle physics
reactions. The (M) and (U) operations then correspond to vertices
(see lower half of Fig.~\ref{fig_diagrams}) which describe
the dual information-conserving processes 
(U)~$2c+\overline{e}\to q$ and (M)~$q+e \to 2c$, summarizing
Eqs.~(\ref{eq_1/4}--\ref{eq_4/4}).
As the $e\overline{e}$ contain no readable information,
they cannot appear in the {\em external} lines of diagrams,
just like the virtual particles of quantum field theory.
\par

We thank Steve Koonin, Asher Peres, Barry Simon,
and Armin Uhl\-mann for enlightening discussions, as well as
an anonymous referee. This work has been funded by NSF grants
PHY 94-12818 and PHY 94-20470, and by
DARPA/ARO through the QUIC Program (\#DAAH04-96-1-3086).

\end{multicols}

\begin{references}
\bibitem{fn_elsewhere} The details of this analysis will be reported elsewhere.
\bibitem{fn_leftover} In this description, the cbits remain unamplified.
The original teleportation scheme can be understood in a similar manner.
Then, the leftover cbits after (U) are in a
product state with $q'$, so that the remaining entropy $S(2c)=2$
should not be included in the entropy balance. It simply originates from the
measurement of two $|0\rangle+|1\rangle$ states in the standard basis,
as revealed by a more detailed analysis.
\end{references}
\end{document}